\documentclass[fleqn,10pt]{SelfArx} 

\usepackage[english]{babel} 

\usepackage{lipsum} 

\definecolor{color1}{RGB}{0,0,90} 
\definecolor{color2}{RGB}{0,20,20} 

\usepackage{hyperref} 
\hypersetup{hidelinks,colorlinks,breaklinks=true,urlcolor=color2,citecolor=color1,linkcolor=color1,bookmarksopen=false,pdftitle={Title},pdfauthor={Author}}

\usepackage{textcomp}

\JournalInfo{\color{white}Journal, Vol. XXI, No. 1, 1-5, 2013} 
\Archive{Additional note} 

\PaperTitle{Diffusion and Perfusion Magnetic Resonance Imaging: Fundamentals and Advances} 

\Authors{Sanam Assili \textsuperscript{1}*} 
\affiliation{\textsuperscript{1}\textit{Medical Physics Department , School of Medicine 
		Tabriz University of Medical Sciences,Tabriz,IRAN,Postal Code: 5166614766}} 
\affiliation{*\textbf{Corresponding author}: assilis@tbzmed.ac.ir} 

\newcommand{\keywordname}{Keywords} 

\Abstract{Over the past few decades, magnetic resonance imaging has been utilized as a powerful imaging modality to evaluate the structure and function of various organs in the human body, such as the brain. Additionally, diffusion and perfusion MR imaging have been increasingly used in neurovascular clinical applications. In diffusion weighted magnetic resonance imaging, the mobility of water molecules is explored in order to obtain information about the microscopic behaviour of the tissues. In contrast, perfusion weighted imaging uses tracers to exploit hemodynamic status, which enables researchers and clinicians to consider this imaging modality as an early biomarker of certain brain diseases. In this review, the fundamentals of physics for diffusion and perfusion MR imaging – both of which are highly sensitive to microenvironmental alterations at the cellular level – as well as their application in the treatment of aging, Alzheimer\textquotesingle s disease, brain tumors and cerebral ischemic injury were discussed. }

\begin{document}

\flushbottom 
\maketitle 
\tableofcontents 
\thispagestyle{empty} 


\section*{Introduction} 

\addcontentsline{toc}{section}{Introduction} 

Various MR techniques have been developed to identify the brain mechanisms underlying complex cognitive, motor, and other behavioral functioning \cite{luypaert2001diffusion}. While diffusion weighted magnetic resonance imaging (DWI) makes use of the translational mobility of water molecules and provides information regarding the microscopic behaviour of the tissues in the presence of macromolecules, the presence and permeability of membranes, and the equilibrium of intracellular-extracellular water, perfusion weighted imaging uses the endogenous and exogenous tracers and obtains hemodynamic status. Research has shown that the use of both modalities is extremely beneficial in the early detection and assessment of various brain issues, such as stroke, tumor characterisation, and the evaluation of neurodegenerative diseases (aging and Alzheimer\textquotesingle s disease) \cite{sullivan2006diffusion}. MR diffusion weighted imaging (DWI) measures the signal loss associated with the random thermal motion of water molecules in the magnetic field to estimate a parameter which is referred to as the apparent diffusion coefficient. This parameter directly defines the translational mobility of the water molecules in the tissues. As mentioned previously, these techniques are used in the early detection and assessment of stroke, tumor characterisation, and the evaluation of multiple sclerosis as well as age-related disease. Perfusion weighted imaging (PWI) allows for the inference of how blood traverses the brain\textquotesingle s vasculature. PWI measures the effects of endogenous or exogenous tracers on MR images in order to derive various hemodynamic quantities, such as cerebral blood volume, cerebral blood flow and mean transit time. Although DWI and PWI can separately provide information, in practice, a combination of both modalities can prove much more useful than a single modality \cite{sullivan2006diffusion}.


\section{Diffusion and Perfusion MRI: Basic Physics}
\subsection{Diffusion weighted imaging}
In MR spectroscopy, signal attenuation is caused by molecular diffusion in the presence of a magnetic field gradient. In 1965, Stejskal and Tanner developed the pulsed gradient, forming the basis of modern diffusion weighted imaging techniques \cite{sullivan2006diffusion}. Brownian motion results in molecular diffusion that represents a constant random flow of individual molecules in a fluid state due to thermal agitation. The average displacement of the molecules remains zero, but over time, the probability of finding an individual molecule at a distance from its point of origin is not zero.  In reality, the root-mean-square displacement increases in proportion to the square root of time, the constant of proportionality being a diffusion constant D, characterising the fluid studied. The diffusion coefficient of pure water is about ${2.2 \times  10^{3} }$ ${\frac{ mm^{2} }{s} }$ Soft tissues tend to behave much like aqueous protein solutions. Additionally, due to the reduced mobility of the water molecules, the corresponding diffusion coefficient is generally smaller than that of pure water. In many tissues, boundaries with various degrees of permeability hinder the free diffusion of water, further decreasing the diffusion coefficient. Applying the Brownian motion model in these circumstances leads to an ‘apparent diffusion coefficient,\textquotesingle  or ADC, which must be distinguished from the diffusion coefficient of free water molecules. In tissues such as white brain matter, an additional complication arises from the fact that molecular mobility is not the same in all directions, i.e., the diffusion process is anisotropic, and the scalar diffusion coefficient must be replaced by a tensor quantity.
Scalar diffusion model
In an isotropic environment, molecular mobility is described by a scalar diffusion coefficient in which the Brownian motion is similar in all spatial directions. In the absence of magnetic field gradients, the signal is unaffected by the presence of incoherent motion. As soon as field gradients are switched on during any stage of the signal preparation, the motion leads to spin dephasing that, due to the random nature of the successive trajectories of each individual molecule, cannot be undone. The result is an exponential attenuation of the original signal, as shown in Equation \ref{eq1}.
\begin{equation}
\label{eq1}
S= S_{0} (N(H),T_{1},T_{2}) x^{-bD}
\end{equation}
Where D is the (apparent) diffusion coefficient of the medium and b is a scalar reflecting the properties of the gradient G (t).

\subsection{Tensor diffusion model}
Based on the morphology, many tissues demonstrate anisotropic diffusion behavior in which the ADC values that are measured using the Stejskal–Tanner sequences are determined by the direction of the sensitizing gradient. For an anisotropic diffusion process, Equation 1 is updated as the following (Equation \ref{eq2}):
\begin{equation}
\label{eq2}
S= S_{0} (N(H),T_{1},T_{2}) x^{- \sum b_{ij}D_{ij}} 
\end{equation}
Where ${i}$ and ${j}$ can be any of the three spatial directions: ${x, y, z}$ in an orthogonal frame of reference. The ${b_{ij}}$ factors characterize the sensitizing gradients along the ${i}$ and ${j}$ directions. This tensor is symmetrical and contains only six independent elements, the determination of which requires the acquisition of images with at least two different diffusion weightings for each of at least six independent directions of the sensitizing gradient. The information in the diffusion tensor may be conceptualized using a “diffusion ellipsoid” picture: the portion of space within which we can expect a molecule to end up due to its aleatory motion, which expands around the point of origin as time goes by and, in general, has the shape of a flattened cigar, reflecting anisotropic mobility.
\subsection{Perfusion weighted imaging}
\begin{figure}[h!]
	\includegraphics[width=\linewidth]{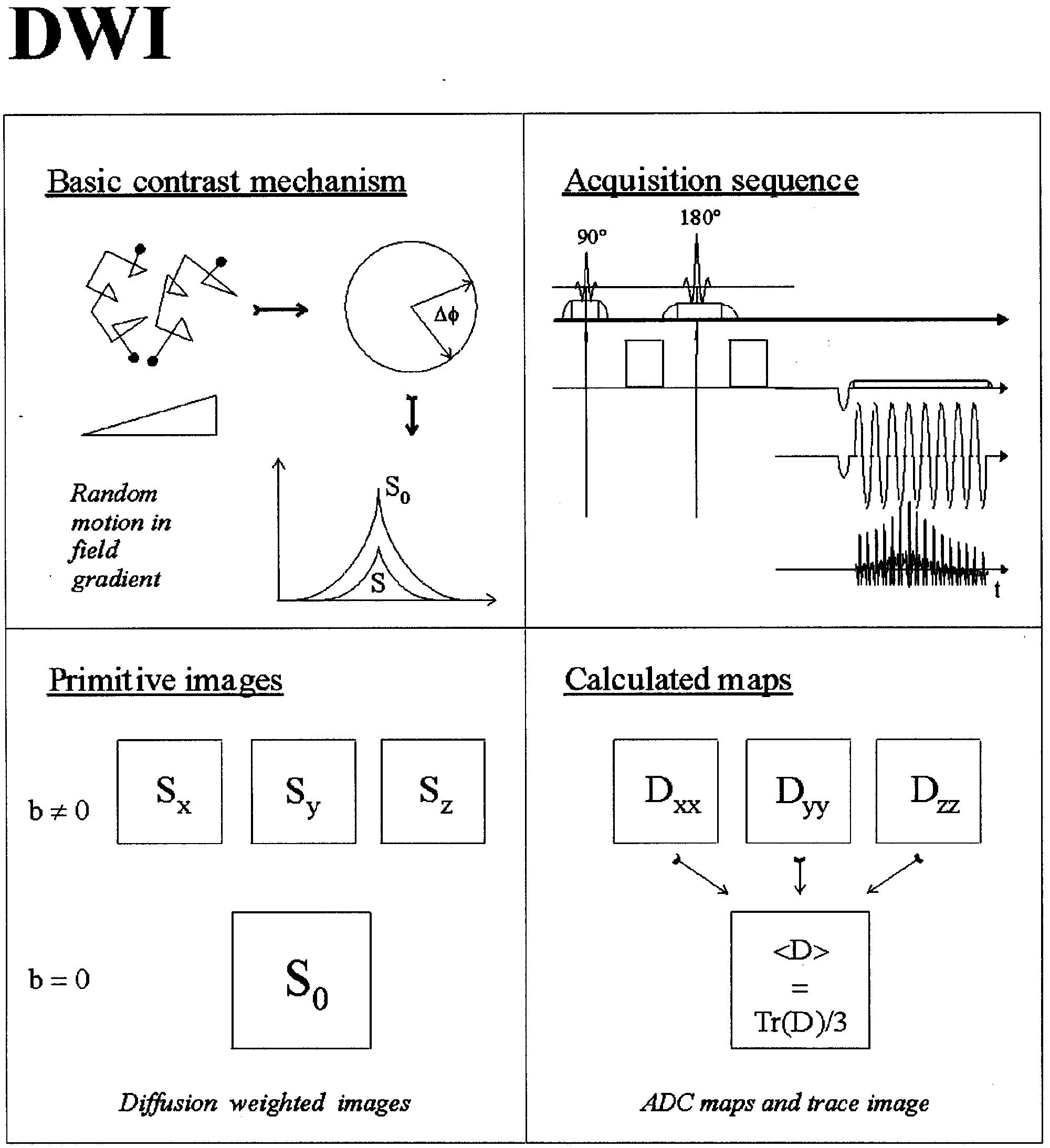}
	\caption{Schematic overview of diffusion weighted MRI methodology (Image courtesy: Luypaert et al.)}
	\label{fig001}
\end{figure}

One of the early approaches to perfusion weighted MRI was to compare the aleatory nature of the motion of blood through the randomly oriented capillaries to that of Brownian motion. In this method, there was success in using the principles of diffusion weighted MR for estimating blood flow. This method, which is also called the “intra-voxel incoherent motion” (IVIM) technique, has recently been replaced by methods that rely on magnetic susceptibility and inflow effects. Susceptibility PWI is based on the passage of intravascular tracers like Gd-DTPA through the capillaries, producing a transient signal loss due to susceptibility effects and allowing first-pass kinetics of the agent to be applied. PWI with arterial spin tagging utilizes the blood itself, with suitably prepared magnetization, as an endogenous tracer. Based on certain assumptions, perfusion imaging allows for the estimation of several important hemodynamic parameters such as cerebral blood volume (CBV), defined as the fraction of the total tissue volume within a voxel occupied by blood; and cerebral blood flow (CBF) or perfusion, defined as the volume of arterial blood delivered to the tissue per minute per tissue volume and the mean transit time (MTT), which corresponds to the average time it takes a tracer molecule to pass through the tissue \cite{sullivan2006diffusion}. Figure \ref{fig001} shows a schematic of a diffusion weighted MRI technique (Figure \ref{fig001}).

\section{Diffusion and Perfusion MRI of the Brain}

\subsection{Aging }

\begin{figure}[h!]
	\includegraphics[width=\linewidth]{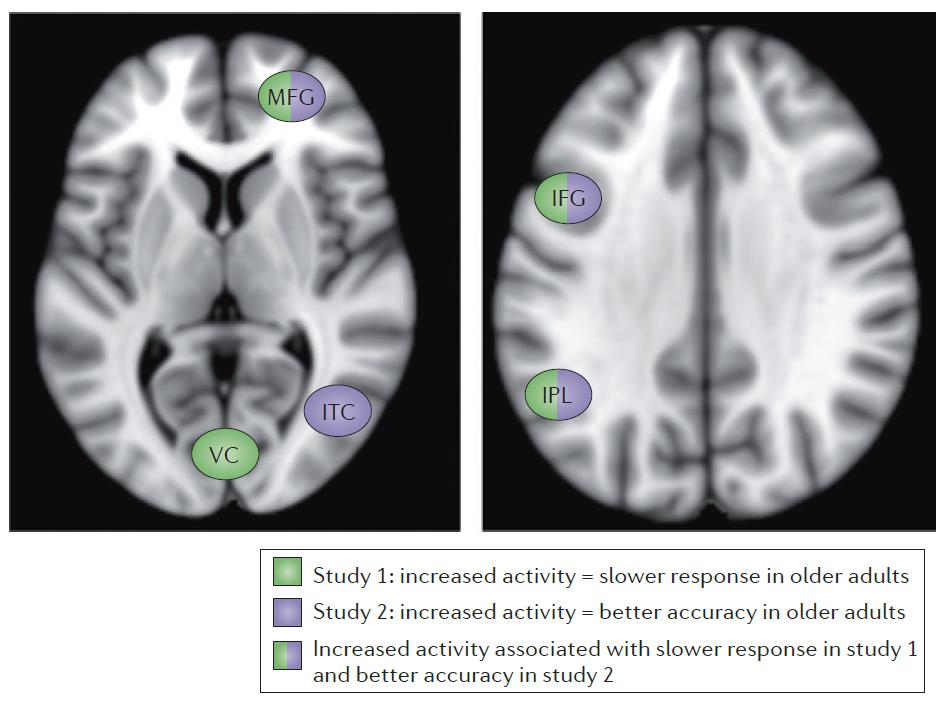}
	\caption{Increased brain activity in older adults may be associated with better or worse task performance (Image Courtesy: Cheryl Grady - Nature)}
	\label{fig002}
\end{figure}

For several years, research has demonstrated that brain structure and functional change by aging and older adults causes particular difficulties with episodic memory, which is defined as the conscious recollection of events. Older adults are also more susceptible to the effects of distracting interference during cognitive tasks and have a generally slower processing speed. Nevertheless, some aspects of cognition are maintained with age, such as semantic memory (the accumulation of knowledge about the world) and emotional regulation. In addition, age differences in cognition are not immutable: for example, the experimental conditions under which memory is studied in older adults can be modified so that age differences are reduced or eliminated. Decreased brain activity has typically been interpreted as a reflection of cognitive deficits in older adults, and increased activity has often been interpreted as compensatory. Another issue is how brain activity is related to other aspects of brain aging, such as changes in structure (volumes or white matter myelination, for example) or changes in neurotransmitters. There is also the question of how findings regarding age differences in brain function might be affected by undetected neuropathological changes due to dementing illnesses. That is, some otherwise healthy older adults might eventually be diagnosed with Alzheimer\textquotesingle s disease, and the ‘silent\textquotesingle  pathological processes in their brains might account for some of the findings tied to age differences reported in the literature \cite{grady2012cognitive}. Normal aging and senescence are defined by various well-established structural magnetic resonance imaging correlates. Enlargement of perivascular spaces and generalized atrophy with enlargement of the ventriculi and the sulci are commonly observed. Deep and subcortical white matter hyperintensities and periventricular hyperintensities are assumed to be due to axonal loss, demyelinization, neuropil atrophy, and vascular damage \cite{minati2007mr}. Figure \ref{fig002} summarizes the results of two studies that differ in terms of how increased brain activity in older adults is associated with task performance. In one of these studies (study 1), the regions shown in green demonstrate a correlation between more activity and slower reaction times on perceptual and working memory tasks in older adults. In another experiment (study 2), several brain regions (indicated in purple) demonstrate a correlation between more activity and more accurate performance in a go/no-go task requiring inhibition of responses. Note that some regions (shown in purple and green) demonstrate an association with better performance in study 2 and the opposite effect in study 1. This discrepancy highlights the complexity of attempts to relate brain activity in older adults to their behaviour, and it indicates that specific relationships between regional brain activity and task performance in older adults are dependent upon the task demands or on the behavioural measure that is assessed (or both). IFG stands for inferior frontal gyrus; IPL, inferior parietal lobe; ITC, inferior temporal cortex; MFG, middle frontal gyrus; and VC, visual cortex.
\begin{figure}[h!]
	\includegraphics[width=\linewidth]{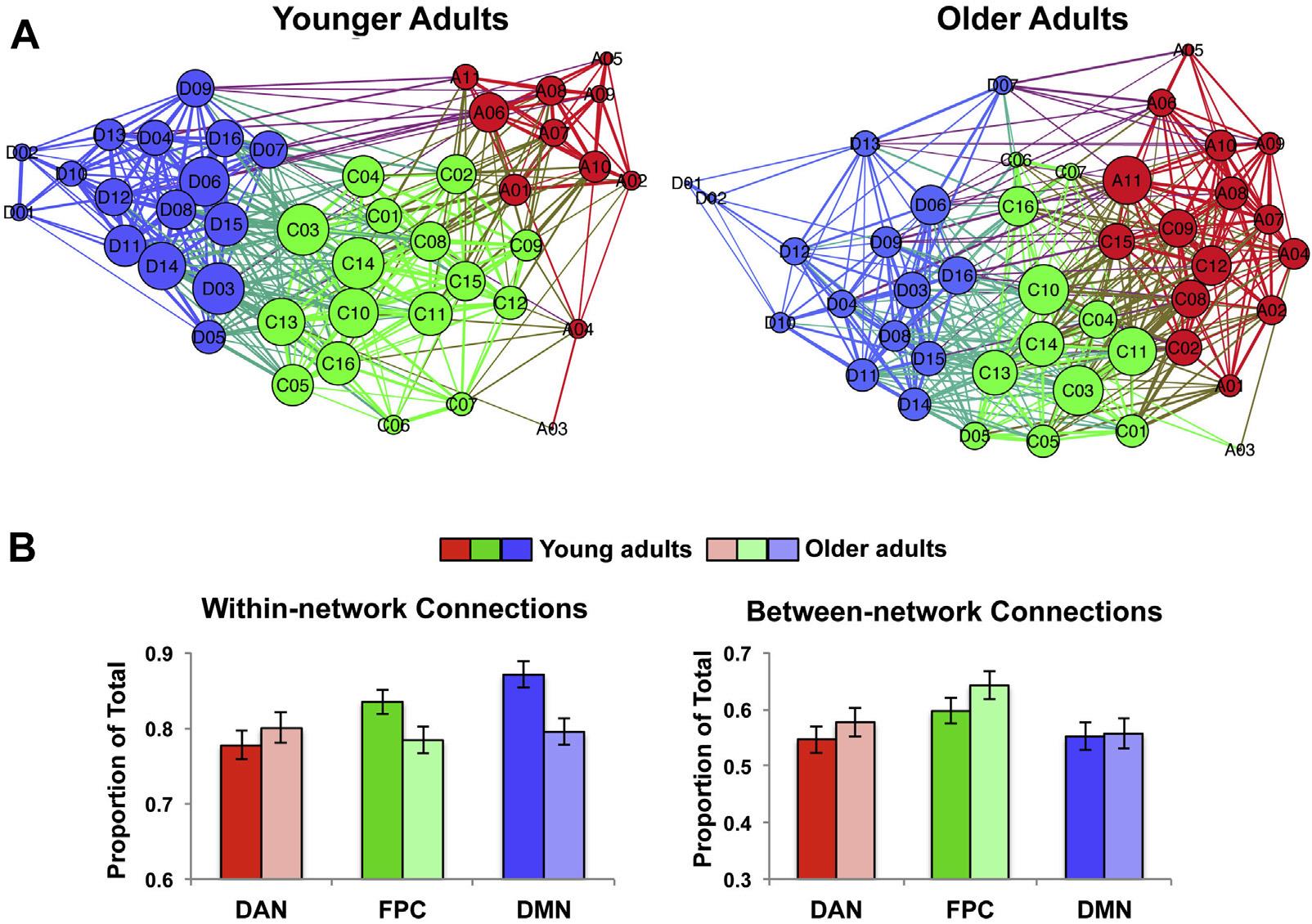}
	\caption{The graphs (A) and within- and between-network connections (B) for young and older adults are shown for the resting state fMRI data. The DAN is represented in red, FPC in green, and DMN in blue. The FPC lies between the other 2 networks in both groups, although some nodes have been assigned to non-canonical networks in the older group. Each metric (in B) was calculated as the proportion of connections within or between networks (out of the total possible) for each node and then averaged across nodes for each network. Reassigned nodes in older adults were considered to be part of the network as they appear in (A) (e.g., nodes C02, C08, C09, C12 and C15 were included in the DAN for the calculations seen in [B]). Abbreviations: DAN, dorsal attention network; DMN, default mode network \cite{grady2016age} \cite{sarraf2016functional} \cite{Sarraf_2016}.}
	\label{fig003}
\end{figure}

\subsection{White matter systems and age-related decline in frontally based functions}

The constellation of frontally based functions, including working memory, set shifting, problem solving, attention, and other executive functions is especially vulnerable to the undesirable effects of advancing age. Although controversy exists regarding the primacy of the frontal lobe hypothesis to explain age-related functional declines, substantial evidence supports the position that aging disproportionately affects the frontal lobe structure. Identification of brain mechanisms responsible for this time-linked functional demise currently focuses on the health of white matter systems necessary to form neural circuits of frontal processing sites. Although performance of executive functions becomes inefficient with age, executive functioning is still possible and in fact may be enhanced when older individuals recruit wider-spread brain systems than those invoked by younger adults. In addition to providing interhemispheric communication for sensory and motor integration, corpus callosum integrity may influence the allocation of resources when attentional or memory capacity is limited. Some have interpreted the bilateral pattern as evidence of decline in functional hemispheric laterality, arising from compromise of the callosum\textquotesingle s capacity to inhibit participation of the hemisphere less suited to accomplish a task. The outcome can appear as \textquotedblleft dedifferentiation\textquotedblright of hemispheric function \cite{luypaert2001diffusion}. Cross-sectional and longitudinal structural MRI studies consistently show that age-related volume increases in CSF-filled spaces which occur primarily at the expense of cortical gray matter, with most showing little volume change in white matter, as shown in Figure \ref{fig003}.

\begin{figure}[h!]
	\includegraphics[width=\linewidth]{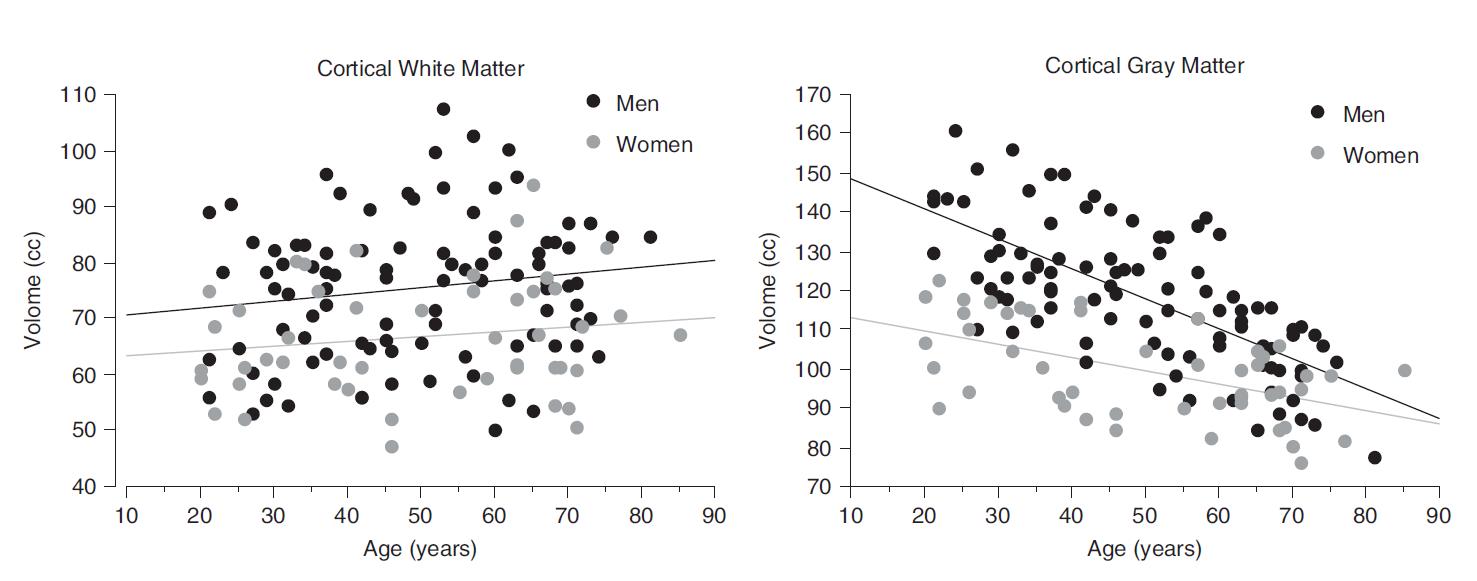}
	\caption{A cross-sectional study revealed age-related decline in cortical gray but not white matter volumes in healthy adults (95 men and 48 women) \cite{luypaert2001diffusion}.}
	\label{fig004}
\end{figure}

\subsection{Imaging Correlates of Aging and Alzheimer\textquotesingle s disease}

\begin{figure}[h!]
	\includegraphics[width=\linewidth]{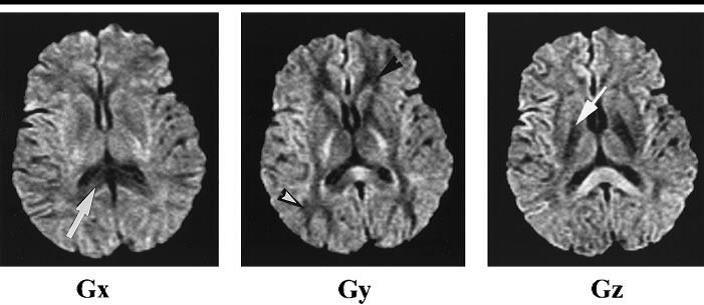}
	\caption{Anisotropic nature of diffusion in the brain. Transverse DW MR images (b = 1,000 sec/mm2); effective gradient, 14 mT/m; repetition time, 7,500 msec; minimum echo time; matrix, 128 3 128; field of view, 200 3 200 mm; section thickness, 6 mm with 1-mm gap) with the diffusion gradients applied along the x (Gx, left), y (Gy, middle), and z (Gz, right) axes demonstrate anisotropy. The signal intensity decreases when the white matter tracts run in the same direction as the DW gradient because water protons move preferentially in this direction. Note that the corpus callosum (arrow on left image) is hypo-intense when the gradient is applied in the x (right-to-left) direction, the frontal and posterior white matter (arrowheads) are hypo-intense when the gradient is applied in the y (anterior-to-posterior) direction, and the corticospinal tracts (arrow on right image) are hypo-intense when the gradient is applied in the z (superior-to-inferior) direction.}
	\label{fig005}
\end{figure}

Alzheimer\textquotesingle s disease and neurodegenerative dementias are growing health problems, since populations are constantly aging. The diagnosis of neurodegenerative diseases is therefore a challenge for modern neuroimaging techniques that extends well beyond their traditional role of excluding other conditions such as neurosurgical lesions \cite{lehericy2007magnetic} \cite{sarraf2014brain} \cite{sarraf2016robust}. A robust  imaging-based algorithm such as Deep Learning, which is able to classify Alzheimer\textquotesingle s disease, will assist scientists and clinicians in diagnosing this brain disorder and will also aid in the accurate and timely diagnosis of Alzheimer\textquotesingle s patients \cite{sarraf2016deep} \cite{sarraf2016deepad}. Additionally, through the assessment of cerebral blood flow in AD with spin labeling techniques, MRI has the potential to assist in the diagnosis of neurodegenerative dementias. In general, structural MRI is used to recognize Alzheimer\textquotesingle s disease, and functional MRI (fMRI) is also used for research purposes. Because this form of brain degeneration affects the brain structure, structurally based MRI techniques are utilized for early detection of Alzheimer\textquotesingle s disease.  In addition to MR morphometry, several recent imaging techniques have demonstrated the potential to detect early abnormalities in AD patients that may be useful in the clinical diagnosis of dementias. These techniques include perfusion, diffusion, spectroscopy and microscopy. Perfusion MRI techniques can assess cerebral perfusion (cerebral blood volume and cerebral blood flow). They therefore have the potential to depict functional deficiencies in a manner that is similar to HMPAO SPECT. Techniques using radiotracers have consistently shown a reduction in cerebral blood flow in patients with AD in the temporo-parietal association areas, the posterior cingulate cortex, and, to a lesser extent, in frontal association areas, while the primary motor and sensory areas are generally spared. Cerebral perfusion can be assessed using contrast-enhanced techniques that require the injection of a paramagnetic contrast agent. In Alzheimer\textquotesingle s disease patients, reduced perfusion was reported in parietal areas using contrast-enhanced perfusion imaging \cite{lehericy2007magnetic}. As mentioned previously, diffusion weighted imaging (DWI) is sensitive to the random motion of water molecules in the brain. Measures of water diffusivity from DWI provide an estimate of the microstructural integrity of the brain parenchyma Increased diffusivity has been reported in the temporal lobe and the posterior white matter of patients with Alzheimer\textquotesingle s disease, as well as MCI subjects. Increased diffusivity may reflect a change in the ultrastructural organization of brain tissue or may represent an index of neuronal atrophy. In contrast, brain anisotropy was decreased in several white matter regions, including the corpus callosum, the cingulum, the superior longitudinal fasciculus or the area of the perforant pathway, reflecting a change in the white matter bundles connecting the affected brain areas in AD patients. In MCI subjects, higher baseline hippocampal diffusivity was associated with a greater risk of progression to AD. Much like voxel-based morphometry, voxel-based comparison of anisotropy maps showed reduced FA in posterior brain regions. DTI abnormalities revealed some correlation with measures of cognitive performance. Left hippocampal volumes were associated with poor verbal memory performance, particularly when associated with high diffusivity values. In Figure \ref{fig005}, the anisotropic nature of diffusion in the brain by transverse DW MR images is demonstrated \cite{schaefer2000diffusion}. 

\subsection{Brain Tumors}

\begin{figure}[h!]
	\includegraphics[width=\linewidth]{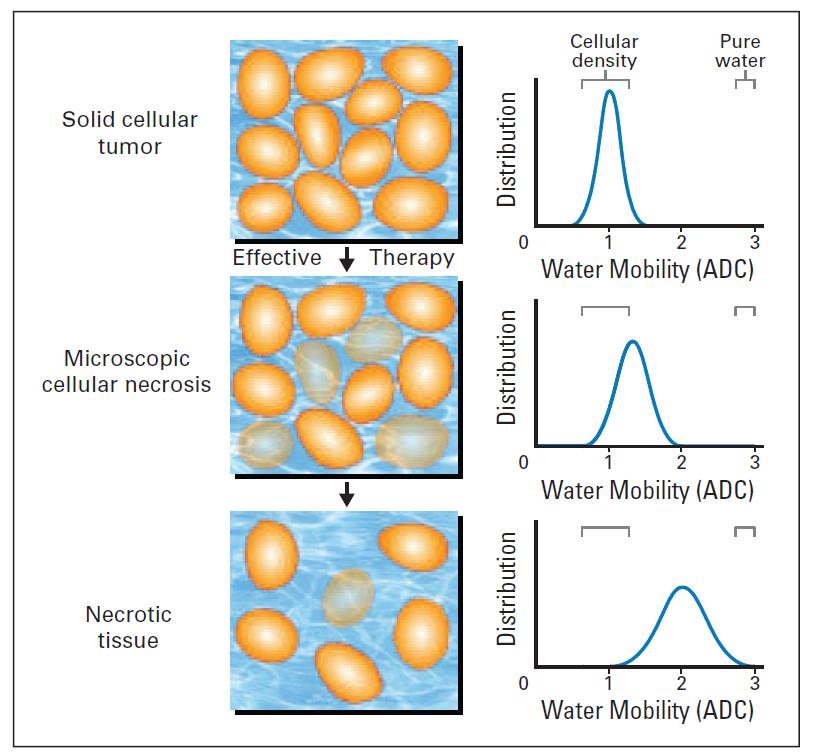}
	\caption{A schematic of the change in cellularity (left) and increased molecular water mobility measured as an apparent diffusion coefficient (ADC; right) as a tumor responds to treatment (top to bottom). For a tumor responding to therapy, an increase in extracellular space/membrane permeability allows for greater water mobility and an increase in the ADC. (Image courtesy: Hamstra et al. 2007)}
	\label{fig006}
\end{figure}

\begin{figure}[h!]
	\includegraphics[width=\linewidth]{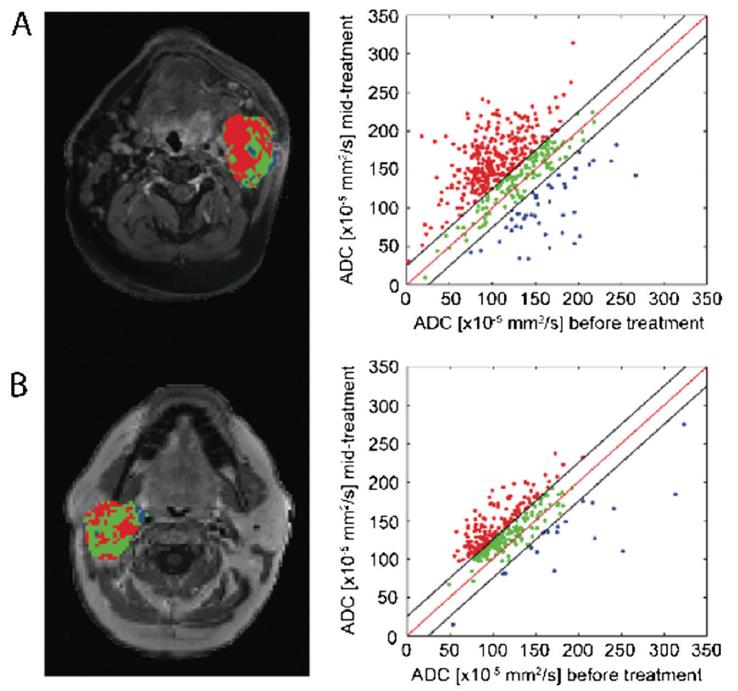}
	\caption{Functional diffusion mapping (fDM) applied to clinical data acquired from patients with head and neck squamous cell carcinoma (HNSCC) diagnosed as pCR (pathological complete response) (A) and PR (partial response) (B). Results from the fDM analysis are presented as color-coded maps superimposed on contrast-enhanced T1-weighted images and scatter plots with axes pre-treatment ADC (x-axis) and post-treatment ADC (y-axis). Color coding is as follows: red, increased ADC values; blue, decreased ADC values; green, unchanged ADC values. (Image courtesy: Galban et al. \cite{galban2009feasibility})}
	\label{fig007}
\end{figure}

Imaging of the response to oncology treatments, in both clinical protocol and standard practice, is a complicated process that has progressed during the last 20 years as modern imaging technologies and newer modalities have been developed. One of the most important questions in the field of oncology is how any tumor response or progression is determined.  A unique method to define tumor response is not completely acceptable. As a result, a new definition for solid tumor responses based on a single linear summation of a small number of target lesions, termed Response Evaluation Criteria in Solid Tumors (RECIST), has been adopted for clinical protocols. This linear summation is both rapid and reproducible, which facilitates its use in clinical trials. RECIST also represents a step forward from previous response criteria, since it takes into account differences in scan thickness, minimum tumor size, and frequency of evaluation. Novel functional imaging modalities such as perfusion MRI, diffusion MRI, [18F] fluorodeoxyglucose positron emission tomography (PET), or PET with newer tracers provide higher resolution and reproducibility of traditional volumetric data. Those imaging modalities now have the need to evaluate potential new biomarkers for tumor response. The main reason for using the imaging modalities mentioned above is to measure the characteristics of the tumor before therapy to determine prognostic value. They also allow for the evaluation of changes in tumors in response to treatment that could function as a surrogate for clinical purposes. Dynamic contrast enhanced MRI, such as diffusion MRI, provides images with high spatial resolution and facilitates the evaluation of tumor vasculature or response to antiangiogenic therapies. Nevertheless, it requires the use of intravenous contrast and relatively complicated post-processing of acquired images \cite{hamstra2007diffusion} \cite{assili2014accurate} \cite{assili2014pharmacokinetic}. The helpfulness of diffusion MRI in the evaluation of stroke led to its exploration as an early marker of tumor response to therapy, and over the last 22 years, similar early diffusion changes have been demonstrated to occur in both preclinical and clinical settings. Given the high concentration of water within biologic tissues, diffusion MRI has focused primarily on measuring the diffusion of protons present within water molecules. Using this method, the movement of water molecules within a cell can be differentiated from that in the extracellular space; however, because extracellular water is better able to diffuse than intracellular water, it is usually the predominant signal in most biologic systems. For instance, diffusion MRI can accurately discriminate between a fluid-filled cyst and a cellular mass that would have more restricted movement of water molecules. Due to a complex interplay of factors in vivo, the actual diffusion coefficient of water cannot be measured directly by MRI; instead, the diffusion coefficient obtained from orthogonal diffusion weighted MRI in all three planes is obtained and is termed the apparent diffusion coefficient (ADC). Diffusion MR measurements are sensitive and can be used to detect and quantify water diffusion values in tissues, which have been proposed to be related to the ratio of intracellular water to extracellular water. Thus, changes in ADC are inversely correlated with changes in cellularity, as shown in Figure \ref{fig006}. In this scenario, increases in ADC will reflect an increase in the mobility of water, either through the loss of membrane integrity or through an increase in the proportion of total extracellular fluid, with a corresponding decrease in cellular size or number, as seen in necrosis or apoptosis. In contrast, decreases in ADC reflect a decrease in free extracellular water, either through an increase in total cellular size or number, as can be seen with tumor progression, fibrosis, or edema.  Given that molecular and cellular changes in response to stress, cytotoxicity, or oxidative injury precede volumetric changes, changes in diffusion MRI have been hypothesized to serve as an early surrogate for later pathologic or radiologic end points. Early preclinical evaluations and studies with rodent brain and breast tumor models revealed that a change in ADC accurately reflected a change in cellularity and could be measured earlier than changes in tumor volume. In animal models, treatment of breast cancer xenografts with cyclophosphamide (150 or 300 mg/kg) produced a significant 30\% to 40\% increase in ADC two days after treatment, which preceded volumetric response as measured using a non-image-based spectroscopic method.
A prospective trial of diffusion MRI in patients with primary brain tumors was initiated at the University of Michigan (Ann Arbor, MI). All patients underwent a baseline scan within one week before the start of treatment, followed by the first intra-treatment scan at three weeks. The majority of patients were treated with a six-week course of fractionated radiation. Therefore, when these patients were assessed for response, less than half of the total course of radiation had been delivered. In an initial report, mean ADC was weakly correlated with subsequent radiographic response. However, fDMs were able to quantify a relatively small responsive mean SEM volume within the tumors of only 8.1\%${\pm}$3.1\% (range, 0\% to 25\%), and the fDM was able to accurately discriminate between patients who had a progressive disease, a stable disease, or a partial response \cite{moffat2005functional} \cite{hamstra2007diffusion} \cite{assili2015automated} \cite{assili2015dynamic}. 
Figure \ref{fig007} shows fDMs [also referred to as parametric response mapping (PRMADC)] with corresponding scatter plots from patients with head and neck squamous cell carcinoma (HNSCC) diagnosed as a complete response (CR) (Fig. 7A) and a partial response (PR) (Fig. 7B) following therapy. Through analysis of the diffusion maps using fDM, heterogeneity in tumor response can be visualized, with red regions denoting a response (i.e., an increase in ADC from baseline) versus stable and decreased ADC regions depicted as green and blue, respectively. As demonstrated in a variety of tumor types, large regions of increased ADC from the baseline (i.e., red voxels) were strongly correlated with treatment response, irrespective of the presence of tumor regions with stable or decreasing ADC values \cite{galban2016diffusion} \cite{galban2009feasibility}.
\subsection{Cerebral Ischemic Injury}

Diffusion magnetic resonance imaging provides an early marker of acute cerebral ischemic injury. Thrombolytic reversal of diffusion abnormalities has not previously been demonstrated in humans. Serial diffusion and perfusion imaging studies were acquired in patients experiencing acute hemispheric cerebral ischemia treated with intra-arterial thrombolytic therapy within six hours of symptom onset. Diffusion weighted imaging (DWI) detects decreases in the self-diffusion of water molecules, appearing as hyper-intensity on DWI sequences associated with a reduced apparent diffusion coefficient (ADC) value. These changes, seen within minutes of ischemia, are related, at least in part, to cellular energy failure and early cytotoxic edema. Perfusion weighted imaging (PWI) provides a qualitative map of relative cerebral blood flow to identify regions of hypo-perfusion. ADC values typically decrease sharply immediately after stroke onset, remain low for at least 72 to 96 hours, then gradually increase, reaching or surpassing normal levels. Serial studies have shown that the typical natural history of early acute ischemic diffusion lesion volumes is to grow over time. In various series, between 62\% and 88\% of patients imaged initially under six hours exhibited lesion growth on follow-up imaging, with the percent change in lesion volume ranging from 32\% to 107\%. It is speculated that this lesion growth may be the result of the gradual failure of energy metabolism in the ischemic penumbra as it is recruited into the infarct core if early reperfusion does not occur. Diffusion/perfusion MRI has been suggested as a means to identify the ischemic penumbra. A prevalent view posits that in humans, the area of diffusion abnormality constitutes an already irreversible core infarction field, and that the penumbra is the region that will show perfusion but not yet diffusion abnormality (diffusion/perfusion mismatch). However, animal studies suggest that this model of the ischemic penumbra may underestimate the volume of tissue that is salvageable early after ischemic onset. When reperfusion occurs within two to three hours in these animal models, the perfusion deficit resolves and is accompanied by partial reversal of DWI and ADC abnormalities. We hypothesize that in humans, as in animals, portions of the DWI and ADC lesions may be salvaged with early reperfusion \cite{kidwell2000thrombolytic}.
\begin{figure}[h!]
	\includegraphics[width=\linewidth]{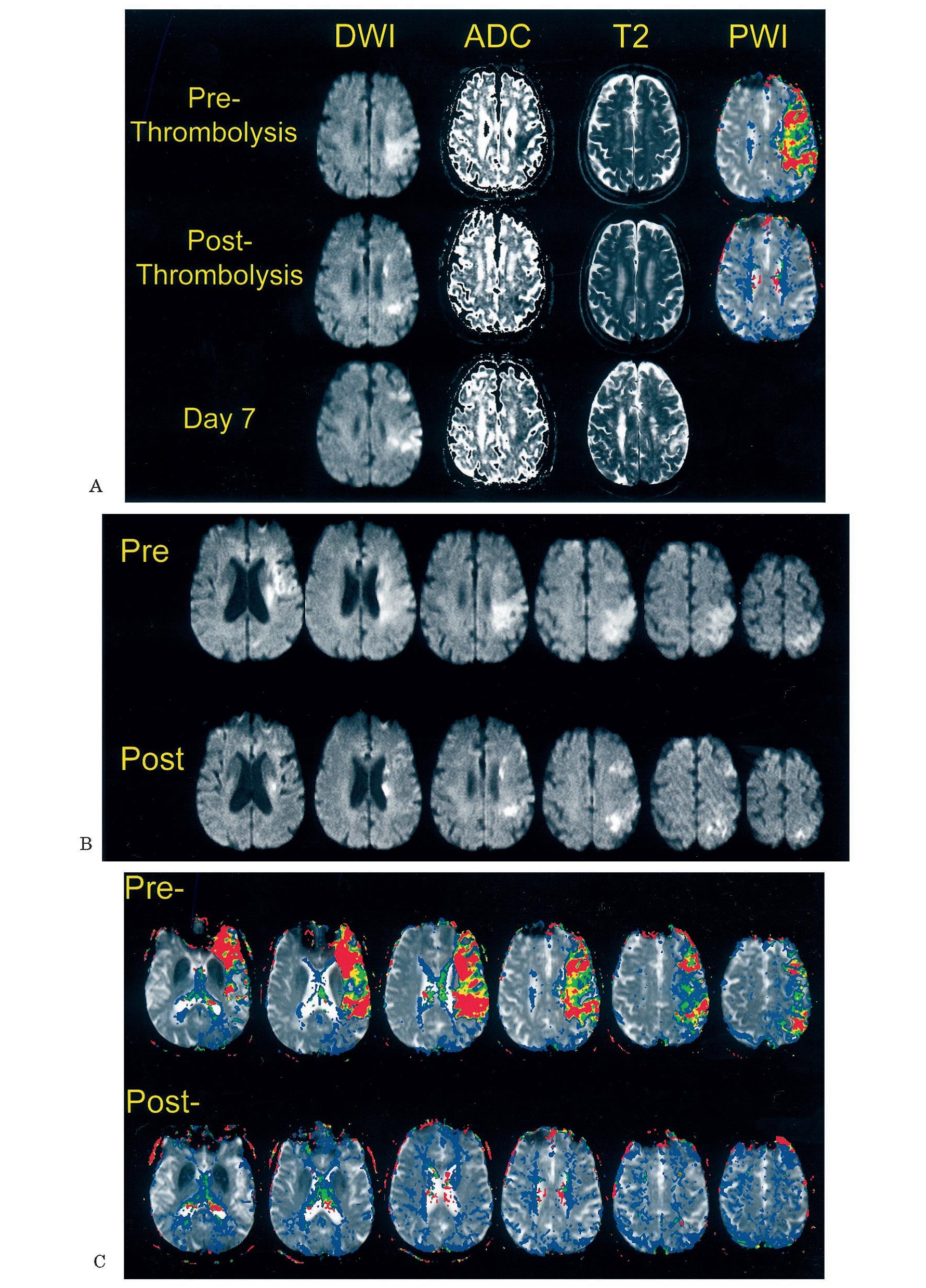}
	\caption{Serial imaging from a 70-year-old man who presented with right hemiparesis and aphasia. The baseline magnetic resonance imaging was obtained 1 hour 15 minutes after symptom onset. (A) Summary display of the baseline, early post-thrombolysis, and day 7 data, including diffusion weighted imaging (DWI), apparent diffusion coefficient (ADC), T2-weighted (T2), and perfusion weighted images (PWI). For the color-coded perfusion weighted images, red represents a bolus delivery delay of greater than 8 seconds, yellow 7 to 8 seconds, green 5 to 6 seconds, and blue 3 to 4 seconds. Early after thrombolysis, diffusion abnormality decreases substantially in size, and the perfusion deficit resolves. The day 7 study shows reduced DWI and ADC lesion volumes and a small final T2 infarct volume. (B) Diffusion weighted images of consecutive axial sections from the same patient at baseline and early after thrombolysis, demonstrating marked reductions in lesion volume at multiple levels. (C) Perfusion weighted images of consecutive axial sections from the same patient at baseline and early after thrombolysis, demonstrating marked reductions in lesion volume at multiple levels. Following thrombolysis, the patient\textquotesingle s aphasia and hemiparesis gradually improved, with National Institutes of Health Stroke Scale score decreasing from 24 at baseline to 17 early after thrombolysis to 13 at day 7. (Image courtesy: Kidwell et al. \cite{kidwell2000thrombolytic})}
	\label{fig008}
\end{figure}
However, one limitation to the widespread use of physiological MRI scanning in guiding early decision making in acute stroke is that diffusion/perfusion MRI is not currently available in many small, community hospitals. Conversely, MRI with echo-planar imaging capability is already widely used in large hospitals and academic stroke centers, where intra-arterial thrombolysis is likely to be most extensively offered, and is rapidly disseminating into the community. Although our series suggests that it is feasible to obtain pre-treatment MRI scans in acute stroke patients, it is important to emphasize that this is a labor-intensive task that requires a well-coordinated stroke team and rapid MRI scanning availability.
\phantomsection
\section{Conclusion}
In this study, the importance as well as the application of diffusion and perfusion MRI were reviewed.  DWI makes use of signal attenuation due to the random motion of water molecules in a strong gradient and allows insight into the microscopic behaviour of the affected tissues, as reflected by the mobility of the water molecules. PWI allows for the assessment of regional cerebral hemodynamics using a variety of methods, among which the first pass endovascular bolus studies are presently the most commonly used in the clinic. These imaging modalities provide crucial information about the aging brain and are considered early biomarkers in the diagnosis of certain brain disorders such as Alzheimer\textquotesingle s disease and brain tumors.  In aging, as an index of brain tissue quality, DTI allows for the examination of regional patterns of neural circuitry degeneration associated with aging that is not possible with other imaging modalities.  Finally, DW MR imaging will also play a greater role in the management of stroke. It may also be helpful in the selection of patients for thrombolysis and in the evaluation of new neuroprotective agents. Additionally, DW MR imaging may prove to be valuable in the evaluation of a wide variety of other disease processes, as described in this review.
\section*{Disclosure statement} 

\addcontentsline{toc}{section}{Disclosure statement} 

The author declares there is no conflict of interest regarding the publication of this manuscript.

\phantomsection
\bibliographystyle{unsrt}


\end{document}